\documentclass{llncs}

\usepackage{microtype}
\usepackage[defaultlines=2,all]{nowidow}
\usepackage{graphicx}
\usepackage{color}
\usepackage[nolist,nohyperlinks]{acronym}
\usepackage{algorithm}
\usepackage{algorithmic}
\usepackage[utf8]{inputenc}
\usepackage{siunitx}
\usepackage{amsmath}
\usepackage{amssymb}
\usepackage[nocompress]{cite}
\usepackage{float}
\usepackage{multirow}
\usepackage[hidelinks]{hyperref}
\usepackage{paralist}
\usepackage{subfigure}
\usepackage{nicefrac}

\begin{document}

\long\def\/*#1*/{}
\newcommand{\todo}[1]{{\color{red}#1}}

\sloppy

\title{Asynchronous Graph Pattern Matching on Multiprocessor Systems}

\author{Alexander Krause, Annett Ungethüm, Thomas Kissinger, Dirk Habich, Wolfgang Lehner\\
       \email{\{firstname.lastname\}@tu-dresden.de}
       \institute{Technische Universität Dresden\\
		Database Systems Group\\
       Dresden, Germany}
		}

\maketitle

\begin{abstract}
Pattern matching on large graphs is the foundation for a variety of application domains.
Strict latency requirements and continuously increasing graph sizes demand the usage of highly parallel in-memory graph processing engines that need to consider non-uniform memory access (NUMA) and concurrency issues to scale up on modern multiprocessor systems.
To tackle these aspects, graph partitioning becomes increasingly important.
Hence, we present a technique to process graph pattern matching on NUMA systems in this paper.
As a scalable pattern matching processing infrastructure, we leverage a data-oriented architecture that preserves data locality and minimizes concurrency-related bottlenecks on NUMA systems.
We show in detail, how graph pattern matching can be asynchronously processed on a multiprocessor system.
\end{abstract}



\section{Introduction}
\label{sec:intro}

Recognizing comprehensive patterns on large graph-structured data is a prerequisite for a variety of application domains such as fraud detection~\cite{Pandit:2007:NFS:1242572.1242600}, biomolecular engineering~\cite{doi:10.1093/nar/28.20.4021}, scientific computing~\cite{DBLP:journals/corr/TasKS17}, or social network analytics~\cite{doi:10.1177/016555150202800601}.
Due to the ever-growing size and complexity of the patterns and underlying graphs, \emph{pattern matching} algorithms need to leverage an increasing amount of available compute resources in parallel to deliver results with an acceptable latency.
Since modern hardware systems feature main memory capacities of several terabytes, state-of-the-art graph processing systems (e.g., Ligra~\cite{DBLP:conf/ppopp/ShunB13}, Galois~\cite{DBLP:conf/sosp/NguyenLP13} or, Green-Marl~\cite{DBLP:conf/asplos/HongCSO12}) store and process graphs entirely in main memory, which significantly improves scalability, because hardware threads are not limited by disk accesses anymore.
To reach such high memory capacities and to provide enough bandwidth for the compute cores, modern servers contain an increasing number of memory domains resulting in a \emph{non-uniform memory access (NUMA)}.
For instance, on a multiprocessor system each processor maintains at least one separate memory domain that is accessible for other processors via a communication network.
Efficient data processing on those systems faces several issues such as the increased latency and the decreased bandwidth when accessing remote memory domains. 
To further scale up on those NUMA systems, pattern matching on graphs needs to carefully consider these issues as well as the limited scalability of synchronization primitives such as atomic instructions~\cite{DBLP:conf/hpdc/YasuiFGBSU16}.  

For efficient \emph{pattern matching} on those NUMA systems, we employ a fine-grained \emph{data-oriented architecture (DORA)} in this paper, which turned out to exhibit a superior scalability behavior on large-scale NUMA systems as shown by Pandis et al.~\cite{Pandis2010} and Kissinger et al.~\cite{Kissinger2014}.
This architecture is characterized by implicitly partitioning data into small partitions that are pinned to a NUMA node to preserve a local memory access.
Since the widely employed \emph{bulk synchronous parallel (BSP)} processing model~\cite{Valiant:1990:BMP:79173.79181}, which is often used for graph processing, does not naturally align with pattern matching algorithms~\cite{6691601}, because a high number of intermediate results is generated that need to materialized and transferred within the communication phase. 
That's why we argue for an asynchronous processing model that neither requires a full materialization nor limits the communication to a distinct global phase.  
Hence, our data partitions are processed by local worker threads that communicate asynchronously via a high-throughput message passing layer to hide the latency of the interconnects between CPUs. 

\noindent\textbf{Contributions.} Following to a discussion of the foundations of graph pattern matching in Section~\ref{sec:patternMatching}, the contributions of the paper are as follows:
\begin{compactenum}[(1)]
	\item{
		We adapt the data-oriented architecture for scale-up graph pattern matching and identify the \emph{partitioning strategy} as well as the design of the \emph{routing table} as the most crucial components within such an infrastructure (Section~\ref{sec:patternOnNuma}).
	}
	\item{
		We describe an asynchronous query processing model for graph pattern matching and present the individual operators a query is composed of.
		Based on the operator characteristics, we identify \emph{redundancy} in terms of partitioning as an additional critical issue for our approach (Section~\ref{sec:queryExecution}). 
	}
	\item{
		We thoroughly evaluate our graph pattern matching approach on multiple graph datasets and queries with regard to scalability on NUMA systems.
		Within our evaluation, we focus on different options for the partitioning strategy, routing table, and redundancy as our key challenges (Section~\ref{sec:evaluation}).
	}
\end{compactenum}

\noindent Finally, we discuss the related work in Section~\ref{sec:relatedWork} and conclude the paper in Section~\ref{sec:conclusion} including promising directions for future work.

\section{Foundations of Graph Pattern Matching}
\label{sec:patternMatching}

Within this paper, we focus on pattern matching for \emph{edge-labeled multigraphs} as a general and widely employed graph data model~\cite{Pandit:2007:NFS:1242572.1242600, doi:10.1177/016555150202800601, doi:10.1093/nar/28.20.4021}.
An edge-labeled multigraph $G(V,E,\rho,\Sigma,\lambda)$ consists of a set of vertices $V$, a set of edges $E$, an incidence function $\rho : E \rightarrow V \times V$, and a labeling function $\lambda : E \rightarrow \Sigma$ that assigns a label to each edge, according to which edge-labeled multigraphs allow any number of labeled edges between a pair of vertices.
A prominent example for edge-labeled multigraphs is RDF~\cite{877487}. 

\begin{figure}[t]
\centering
	\subfigure[b][Example query pattern.]{\label{fig:exQuery}\includegraphics[width=0.17\textwidth]{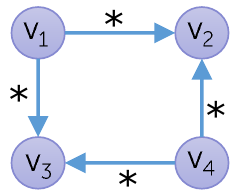}}
~
	\subfigure[b][Example graph.]{\label{fig:exGraph}\includegraphics[width=0.27\textwidth]{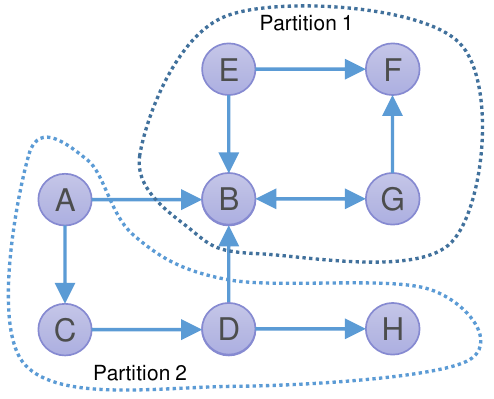}}
~
	\subfigure[b][Derived operators.]{\label{fig:exOps}\includegraphics[width=0.45\textwidth]{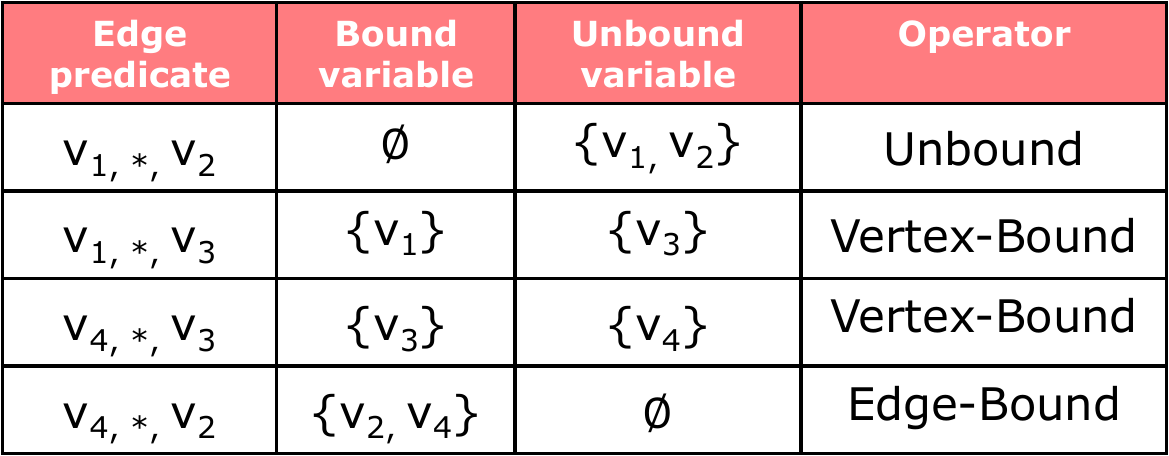}}
\caption{A pattern matching query for an example graph and the resulting query operators.}\label{fig:examples}
\end{figure}

\emph{Pattern matching} is a declarative topology-based querying mechanism where the query is given as a graph-shaped pattern and the result is a set of matching subgraphs~\cite{Tran:2009:TEQ:1546683.1547416}.
For instance, the \emph{query pattern} depicted in Figure~\ref{fig:exQuery} searches for a vertex $V_1$, that has two outgoing edges targeting $V_2$ and $V_3$. 
Additionally, the query pattern seeks a fourth vertex $V_4$ which also has two outgoing edges to the same target vertices. 
The query pattern forms a rectangle with four vertices and four edges of which we search for all matching subgraphs in a graph.
A well-studied mechanism for expressing such query patterns are \emph{conjunctive queries (CQ)}~\cite{Wood:2012:QLG:2206869.2206879}, which decompose the pattern into a set of \emph{edge predicates} each consisting of a pair of vertices and an edge label.
Assuming a wildcard label, the exemplary query pattern in Figure~\ref{fig:exQuery} can be decomposed into the conjunctive query $\{( \pmb{V_1} , *, V_2), ( \pmb{V_1}, *, V_3), ( \pmb{V_4}, *, V_3), ( \pmb{V_4}, *, V_2)\}$, where the bold vertices represent the source vertex of an edge. 
These four edge predicate requests form a sequence, that is processed by starting at each vertex in the data graph, because the query pattern does not specify a specific starting vertex.

For the graph in Figure~\ref{fig:exGraph}, the vertex $A$ is a potential match for $V_1$ of the query pattern.
Following the outgoing edges of $A$, the vertices $B$ are found as a potential match for $V_2$ and $C$ as a potential match for $V_3$.
After matching three of four query vertices, we need to find a vertex, which has an outgoing edge towards $B$ and $C$ to complete the query pattern.
The vertex $D$ has an outgoing edge towards $B$ but no edge targeting the vertex $C$.
Since there is no vertex that has an outgoing edge for $C$, the potential matches are discarded because they can not fulfill the requested edge predicates.
Starting at vertex $E$ performing the same steps again, the system will generate the result $E, F, B, G$ as another match for the query pattern.

\section{Scalable Graph Pattern Matching Architecture}
\label{sec:patternOnNuma}
\begin{figure}[t]
\centering
\includegraphics[width=\textwidth]{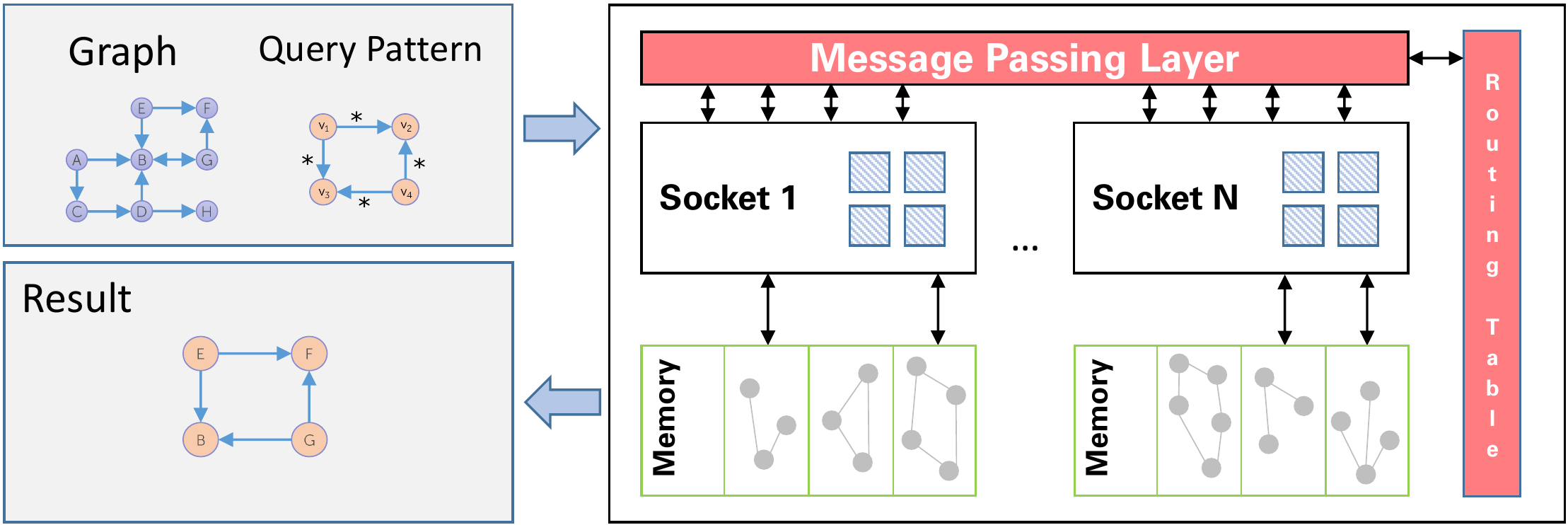}
\caption{Scalable graph pattern matching based on a data-oriented architecture~\cite{Kissinger2014,Pandis2010}.}\label{fig:dora}
\end{figure}

In this section, we describe the general architecture of our graph processing infrastructure and the resulting challenges for \emph{graph pattern matching}.
Since current hardware trends towards an increasing amount of parallelism and main memory capacities, a non-uniform memory access becomes more and more common to allow hardware resources to scale up to such dimensions. 
To address the issue of increasing NUMA effects and to achieve scalability of graph pattern matching algorithms inside of a single machine, we propose an adapted data-oriented architecture for graph processing as depicted on the right hand side of Figure~\ref{fig:dora}. 
Within this architecture, we bind worker threads to each of the available hardware threads of the multiprocessors leading to multiple compute clusters.
The key characteristic of a NUMA system are local main memory regions that belong to specific sockets.
All sockets can access another sockets main memory via internal interconnects.
However, this comes at the cost of reduced bandwidth and increased latency \cite{Kissinger2014}.
Thus, the data-oriented architecture describes the principle of implicitly partitioning the data (e.g., graphs) and storing the individual partitions in the local main memory of a specific socket.
During the processing, each worker threads process local partitions (exclusively locked) that have pending data processing requests.
This design eliminates the necessity of fine-grained locks over data structures and allows all workers to process local data only.

\noindent \textbf{Partitioning Strategy.} However, partitioning a graph will most likely lead to edges, which span over multiple partitions, like the edges $A \rightarrow B$ and $D \rightarrow B$ in Figure~\ref{fig:exGraph}.
For instance, if vertex $A$ is considered as a potential match for a query pattern, the system needs to lookup vertex $B$ in another partition.
Moving to another partition requires that the complete matching state needs to be transferred to another worker, which requires communicational efforts between the two responsible workers.
Our previous example would therefore lead to the transfer of the matching state $\{{(V_1 \rightarrow A), (V_2 \rightarrow ?), (V_3 \rightarrow C), (V_4 \rightarrow ?)}\}$ from the worker of partition 2 to the worker of partition 1.
In best case, both workers reside on the same socket, which means that a local message can be sent.
In worst case, worker 2 needs to send a message to a worker on another socket, which reduces the performance because of higher communication latency.
Hence, the selection of the \emph{partitioning strategy} is crucial when adapting the data-oriented architecture for graph pattern matching, because locality in terms of the graph topology is important~\cite{EuroparPartitioning}.

\noindent \textbf{Routing Table.} Because one partition can not always contain all the necessary information for one query, it is inevitable to communicate intermediate results between workers.
The communication is handled by a high-throughput message passing layer, which hides the latency of the communication network, as depicted in Figure~\ref{fig:dora}.
The system stores the target socket and partition information in a crucial data structure, the \emph{routing table}.
The routing table determines the target partition as well as the target NUMA node per vertex.
Therefore the design of the routing table needs to be carefully considered, because real world graphs often feature millions of vertices and billions of edges.

Since \emph{routing table} and \emph{partitioning strategy} depend on each other, we consider the following three design options for our discussion and evaluation:
\begin{compactenum}
	\item[\textbf{Compute Design.}]{
		The compute design is a combination of a hash function as \emph{routing table} and a locality-agnostic \emph{partitioning strategy}.
		Hash partitioning is easy to compute, because it only needs to consider the id of a vertex to assign it to a partition.
		This implementation calculates the target partition on the fly and does not use any additional data structures.
		Nevertheless, due to the basic routing table, the partitioning strategy can not take any topology-based locality information into account.
}
	\item[\textbf{Lookup Design.}]{
		The lookup design is the opposite to the compute design and is a combination of a hash table -- instead of a function -- as \emph{routing table} and a locality-aware \emph{partitioning strategy}.
		The routing table is represented as a hash map that contains a one-to-one mapping of all vertices of the graph to their respective partitions.
		Thus, we precompute a graph partitioning, which considers the locality of a vertex' neighborhood.
		This approach leads to a routing table, which is as big as the number of vertices, because we need to store the partition for every single vertex in the graph.
		As partitioning strategy, we use the well known \emph{multilevel k-Way} partitioning to create a disjoint set of partitions.
		This heuristical approach creates partitions with high locality and tries to minimize the edge cut of the partitioning \cite{karypis1998fast}.
		}
\end{compactenum}		

\noindent
Both, the \emph{compute} and the \emph{lookup design} face advantages and disadvantages.
On the one hand, the \emph{compute design} is the fastest implementation for a routing table but lacks the ability to consider graph properties like locality or semantic relationships between vertices, to create well balanced locality-aware partitions.
On the other hand, the \emph{lookup design} is able to exploit such graph properties, which comes at the price of an additional storage overhead leading to many remote memory accesses on NUMA systems, because the routing table does not fit into the caches of the individual multiprocessors.
To overcome the disadvantages of both designs, we propose a \emph{hybrid design} that combines the low memory footprint of the \emph{compute design} and the locality awareness of the \emph{lookup design}:  

\begin{compactenum}
	\item[\textbf{Hybrid Design.}]{
		The hybrid design is a combination of a range \emph{routing table} and a locality-aware \emph{partitioning strategy}.
		To enable this combination, we employ a dictionary as auxiliary data structure that maps virtual vertex ids to the original vertex ids.
		This dictionary is only used for converting the virtual ids of the final query results and is generated after the locality-aware partitioning strategy was applied.
		The range-based routing table maps dense ranges of virtual ids to the respective partition and has very low memory footprint such that the routing table easily fits into the cache of the multiprocessors.  
	}
\end{compactenum}

\noindent
Within our evaluation (cf., Section~\ref{sec:evaluation}), we will (1) prove that our adapted data-oriented architecture is able to scale up for graph pattern matching queries on NUMA systems and (2) we will evaluate the impact of the design options for the combination of \emph{routing table} and \emph{partitioning strategy}. 
\section{Graph Pattern Matching Processing Model}
\label{sec:queryExecution}
The architecture introduced in Section~\ref{sec:patternOnNuma} needs specific operators for pattern matching on NUMA systems.
We identified three logical operators, which are necessary to model \emph{conjunctive queries} as described in Section~\ref{sec:patternMatching}:
\begin{compactenum}
	\item[\textbf{Unbound Operator.}]{
	The unbound operator performs a parallel vertex scan over all partitions. 
	There are two types of this operator, based on the edge predicate.
	If the user specifies a wildcard as shown in Figure~\ref{fig:exQuery}, this operator returns all edges between two vertices. 
	By specifying a certain edge predicate, the operator returns only edges with this specific label. 
	The unbound operator is always the first operator in the pattern matching process.
	}
	\item[\textbf{Vertex-Bound Operator.}]{
	The vertex-bound operator takes an intermediate pattern matching result as input and tries to match new vertices in the query pattern. 
	Based on the direction of the requested edge in the query pattern, the input of the vertex-bound operator is either a source or a target vertex.
	}
	\item[\textbf{Edge-Bound Operator.}]{
	The edge-bound operator ensures the existence of additional edge predicates between vertices which are matching candidates for certain vertices of the query pattern. 
	It performs a data lookup with a given source and target vertex as well as a given edge label. 
	If the lookup fails, both vertices are eliminated from the matching candidates.
	Otherwise the matching state is passed to the next operator or is returned as final result.
	}	
\end{compactenum}

\noindent
To actually compose a \emph{query execution plan (QEP)}, the query compiler sequentially iterates over the \emph{edge predicates} of the \emph{conjunctive query}.
For each edge predicate, the query compiler determines whether source and/or target vertex are bound and selects the appropriate operator for the respective edge predicate.
Afterwards, source and target vertex are set to \emph{bound} and the query compiler continues with the next edge predicate.
For the example query pattern in Figure~\ref{fig:exQuery}, the resulting operator assignments of the QEP are shown in Figure~\ref{fig:exOps}.
The first edge predicate is always mapped to the \emph{unbound operator}, because no vertices are initially \emph{bound} and all edges of the graph potentially match this part of the query pattern.
The second and third edge predicate gets mapped to the \emph{vertex-bound operator}, because only one vertex of the predicate is currently bound to a specific vertex.
Finally, the last operator is mapped to the \emph{edge-bound operator}, because source and target vertex are known and thus, bound.

The compiled QEP is executed by invoking the first operator in the sequence.
Due to the nature of the data-oriented architecture, individual operators are processed by worker threads that operate on the respective partitions of the graph.
Each operator is asynchronously processed in parallel and generates new messages that invoke the next operator in the QEP.
Hence,  different worker threads can process different operators of the same query at the same point in time.
Based on the operator and its parametrization, we distinguish two ways of addressing a message that are related to the \emph{routing table}:

\begin{itemize}
	\item[\textbf{Unicast.}] {
		A unicast addresses a single graph partition and requires that the source vertex is known respectively bound by the operator.
		This case occurs for the \emph{vertex-bound operator} if the source vertex is bound and for the \emph{edge-bound} operator.
		The usage of a unicast is desirable, because the respective message only needs to be processed on a single graph partition, which is advantageous in terms of scalability.
	}
	\item[\textbf{Broadcast.}] {
		A broadcast addresses all partitions of a graph, which (1) increases the pressure on the message passing layer and (2) requires the message to be processed on all graph partitions and thus, negatively affects the scalability.
		A broadcast message is initially generated to execute the \emph{unbound operator} that triggers the pattern matching query.
		Additionally, \emph{vertex-bound operators} that bound the target vertex require a broadcast.
	}
\end{itemize}

While the \emph{broadcast} for the initial \emph{unbound operator} is inevitable and is only executed once per query, the broadcasts generated by \emph{vertex-bound operators} significantly hurt the scalability of our approach.
The cause of this problem is inherently given by the data-oriented architecture, because a graph can either be partitioned by the source or the target vertex of the edges.
Hence, we identify \emph{redundancy} in terms of partitioning as an additional challenge for our approach.
To reduce the need for broadcasts to the initial \emph{unbound operator}, we need to redundantly store the graph partitioned by source vertex and partitioned by target vertex.
However, the need for redundancy depends on the query pattern as well as on the graph itself as we will show within our evaluation. 

\section{Evaluation}
\label{sec:evaluation}

\begin{figure}[b]
	\centering
	\subfigure[b][\emph{V} Query]{\label{fig:VQuery}\includegraphics[width=0.3\textwidth]{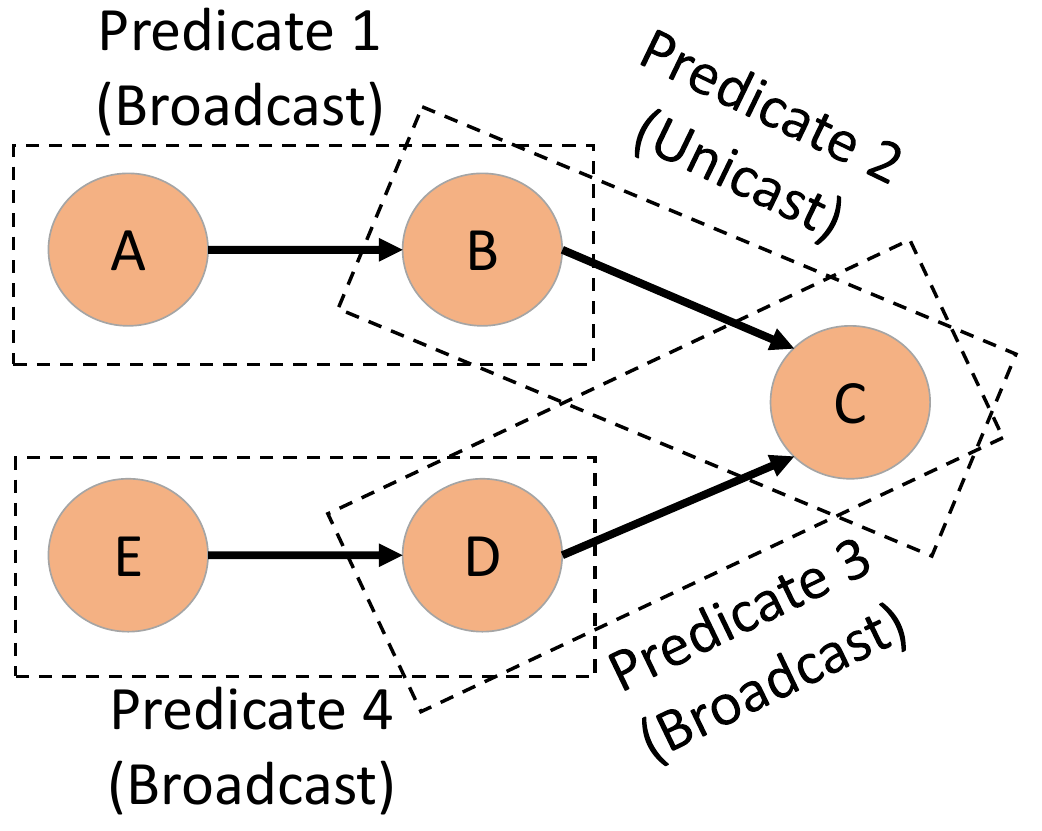}}
	\hfill   
	\subfigure[b][\emph{Quad} Query]{\label{fig:QuadQuery}\includegraphics[width=0.25\textwidth]{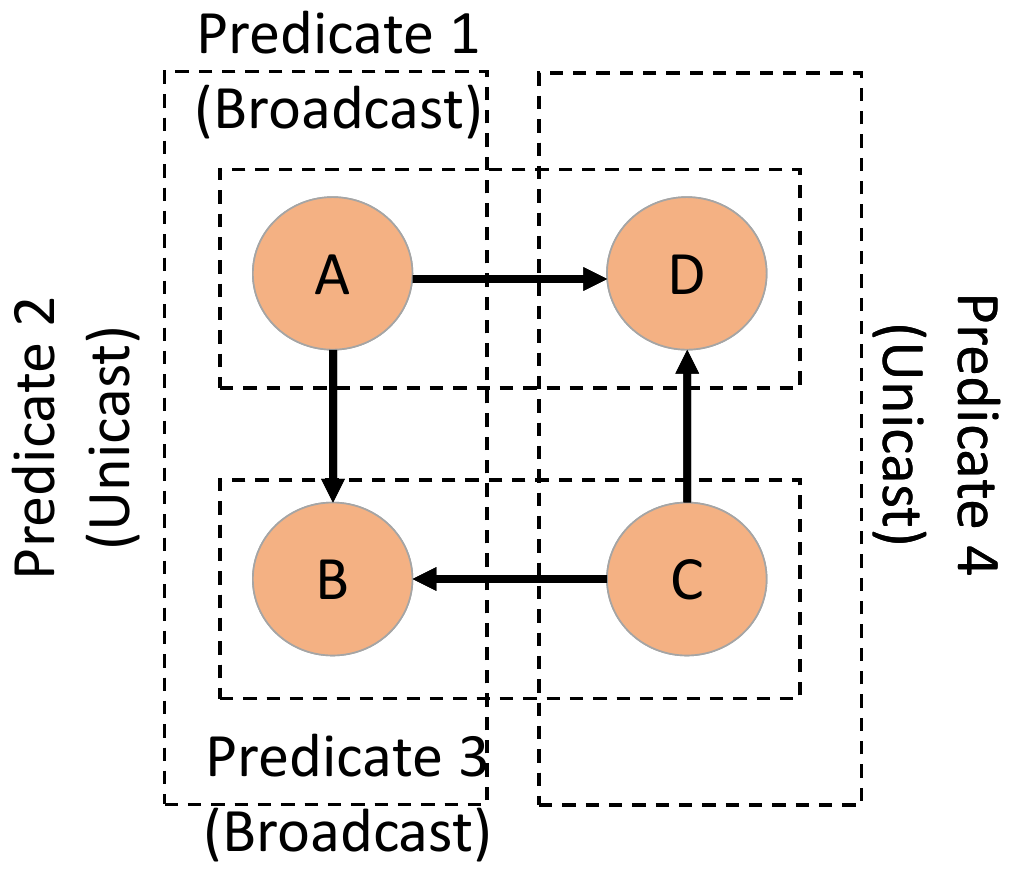}}
	\hfill
	\subfigure[b][Graph meta information]{\label{fig:GraphMeta}\includegraphics[width=0.4\textwidth]{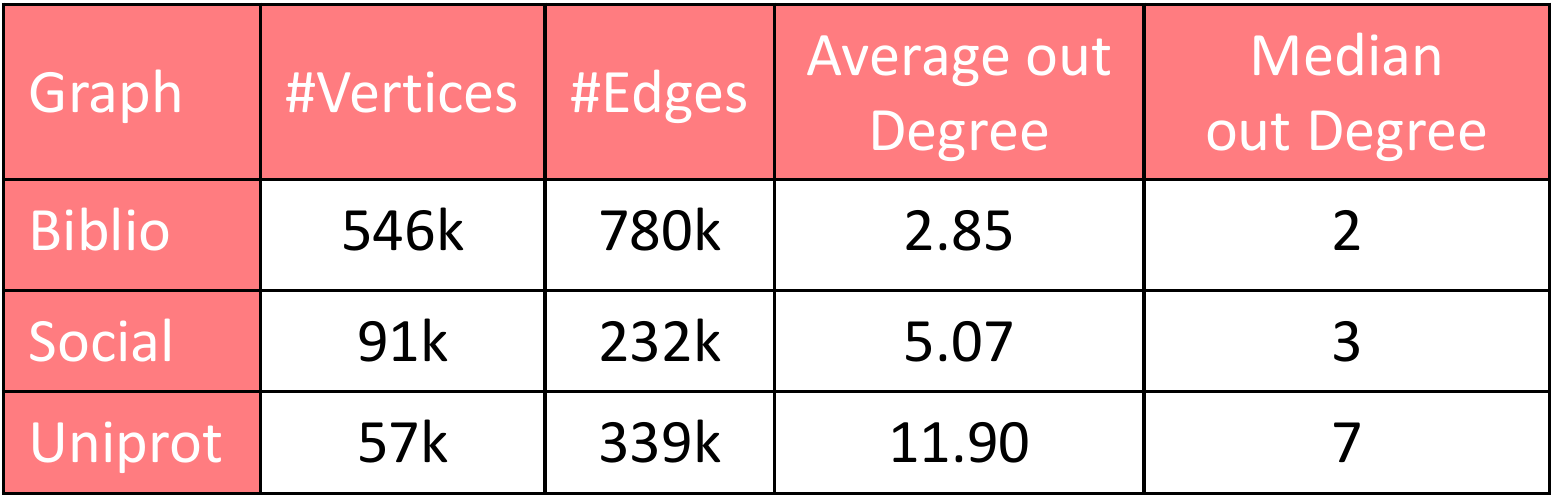}}
	\caption{Query patterns and test graphs for the medium-scale system.}
	\label{fig:Queries}
\end{figure}

Within our evaluation, we investigate the influence of the \emph{routing table} and \emph{partitioning strategy} (cf., Section~\ref{sec:patternOnNuma}) as well as the impact of \emph{redundancy} (cf., Section~\ref{sec:queryExecution}) on the scalability of the \emph{pattern matching} process.
Our NUMA system consists of $4$ sockets each equipped with an Intel Xeon CPU E7-4830 -- resulting in $32$ physical cores and $64$ hardware threads -- and $128$ GB of main memory.
We use three different test data graphs of varying application domains that are generated via the graph benchmark framework gMark~\cite{BBCFLA16}. 
The properties of the graphs are listed in Figure~\ref{fig:GraphMeta}.
Additionally, we defined two \emph{conjunctive queries} as depicted in Figure~\ref{fig:VQuery} and \ref{fig:QuadQuery}:
(a) the \emph{V} query shapes a V with five vertices and four edges and
(b) the \emph{Quad} query is a rectangle, which consists of four vertices and four edges.
For both queries, four \emph{edge predicate} evaluations are necessary.
The evaluation of the edge predicates happens as follows:

\begin{compactenum}
\item[\textbf{V Query.}]{
	The first edge predicate uses the \emph{unbound operator} that is broadcasted to all partitions.
	The intermediate result is a set of target vertices, which are used as source vertices for the second request (\emph{vertex-bound operator}). 
	The intermediate result is a set of destination vertices, which are destination vertices for the third edge predicate  (\emph{vertex-bound operator}). 
	Assuming the absence of \emph{redundancy}, this requests needs to be broadcasted to all partitions, because the source vertex for this edge predicate request is unknown. 
	The same applies for the evaluation of the fourth edge predicate.
}
\item[\textbf{Quad Query.}] {
	The edge predicate evaluation of the \emph{Quad} query can be achieved by employing broadcasts and unicasts in an alternating order as shown in Figure~\ref{fig:exOps} (cf., Section~\ref{sec:queryExecution}).
}
\end{compactenum}

\noindent As the edge predicate evaluation of the two queries suggests, pattern matching is a combination of unicasts and broadcasts within a partitioned environment that depends on the query pattern.
Moreover, the number of generated messages in the respective query stages depends on the found matches and thus, on the graph itself.
In the following, we start with presenting our measurements for the \emph{routing table} and \emph{partitioning strategy} followed by the experimental results for \emph{redundancy} in terms of partitioning.
Finally, we will combine all three aspects and demonstrate their impact on the overall scalability of our approach.

\subsection{Routing Table and Partitioning Strategy}
\label{sec:evaluationRouting}

To investigate the influence of the combinations of \emph{routing table} and the \emph{partitoning strategy} on the query performance, we use the \emph{biblio graph} as described in Figure~\ref{fig:GraphMeta} with a scale factor of up to 32 and the \emph{V query} as pattern.
Since each worker needs to access the routing table during the pattern matching execution, we additionally measured the average time spent in the \emph{routing table} per worker.

\begin{figure}[t]
	\centering
	\subfigure[b][Routing time]{\label{fig:routingTime_nored}\includegraphics[width=0.48\textwidth]{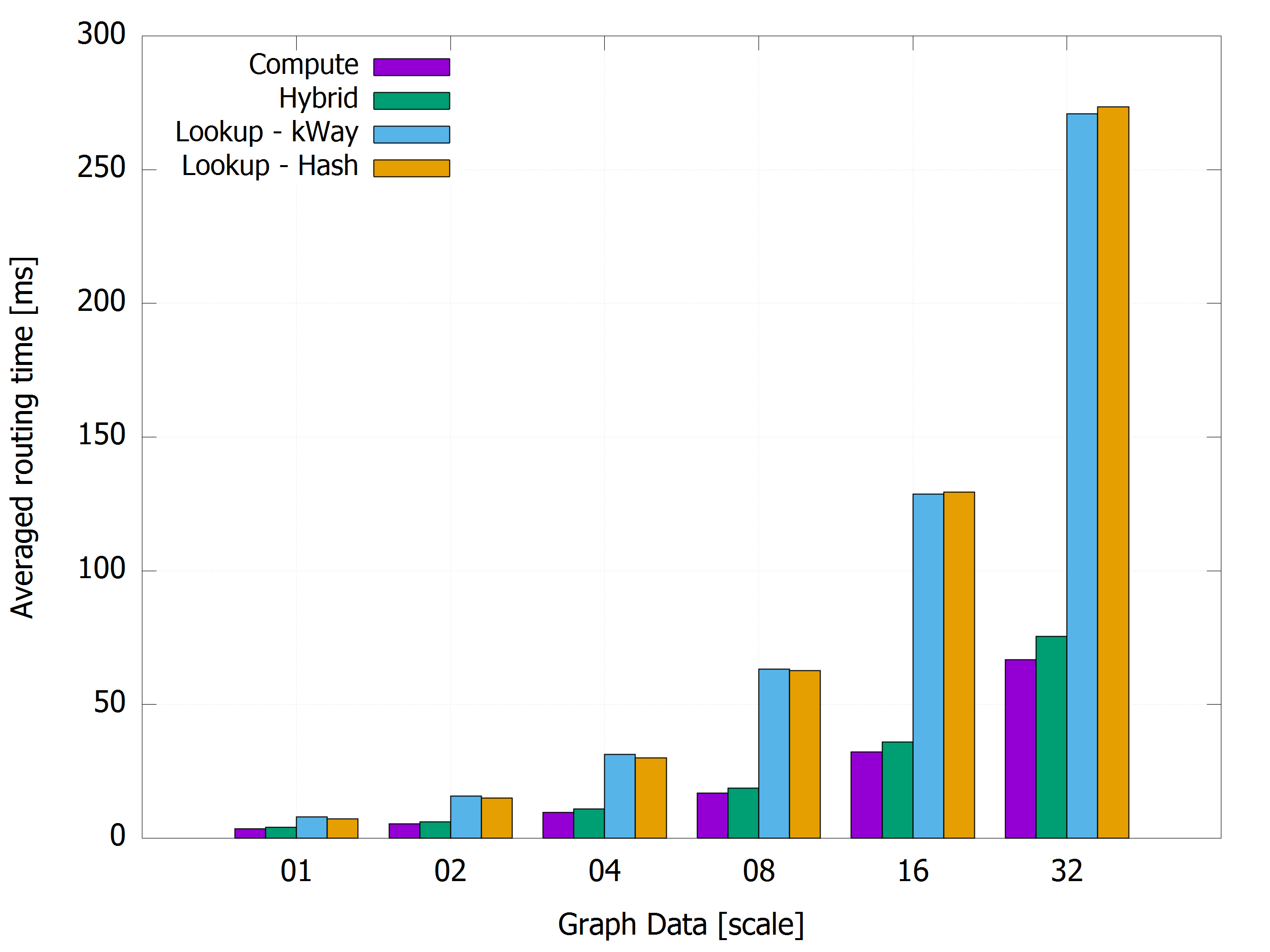}}
	\hfill   
	\subfigure[b][Query runtime]{\label{fig:queryTime_nored}\includegraphics[width=0.48\textwidth]{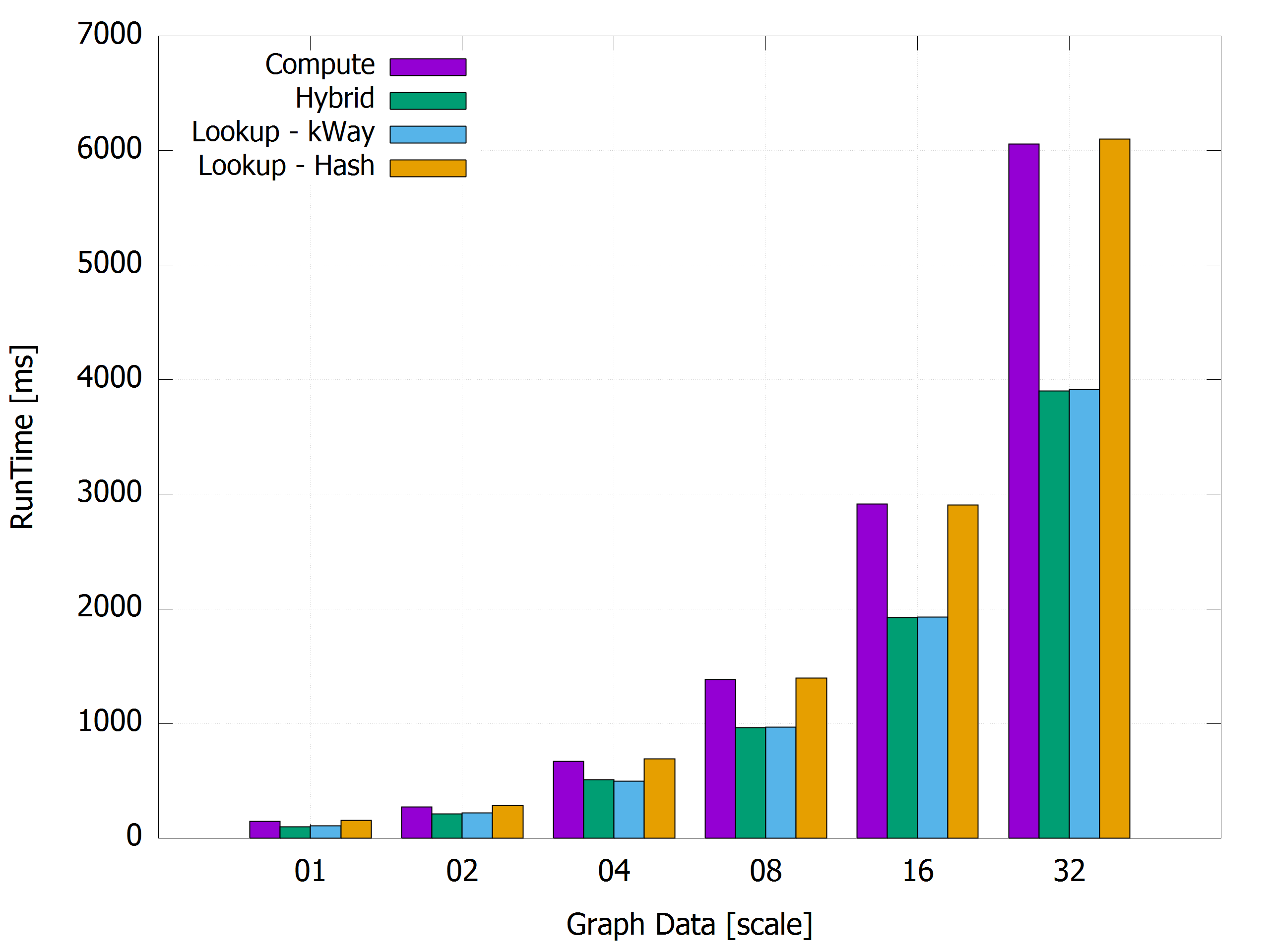}}
	\caption{\emph{V} query on the Biblio graph using different scale factors.}
	\label{fig:routingVsQuerytime_nored}
\end{figure}
~
\begin{figure}[t]
	\centering
	\subfigure[b][\emph{Compute} design]{\label{fig:detailed_hv}\includegraphics[width=0.48\textwidth]{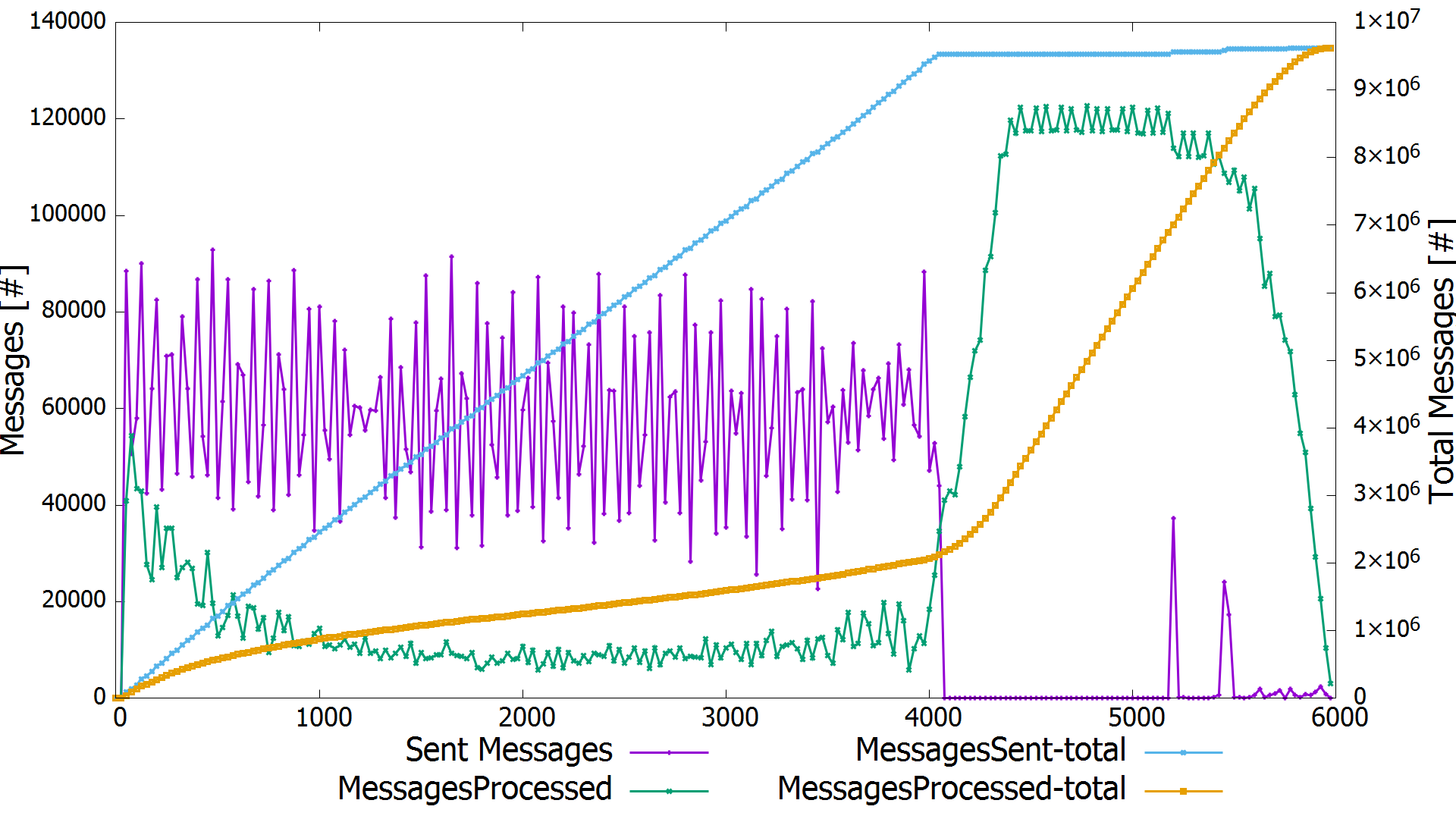}}
	\hfill   
	\subfigure[b][\emph{Lookup-Hash} design]{\label{fig:detailed_hvwt}\includegraphics[width=0.48\textwidth]{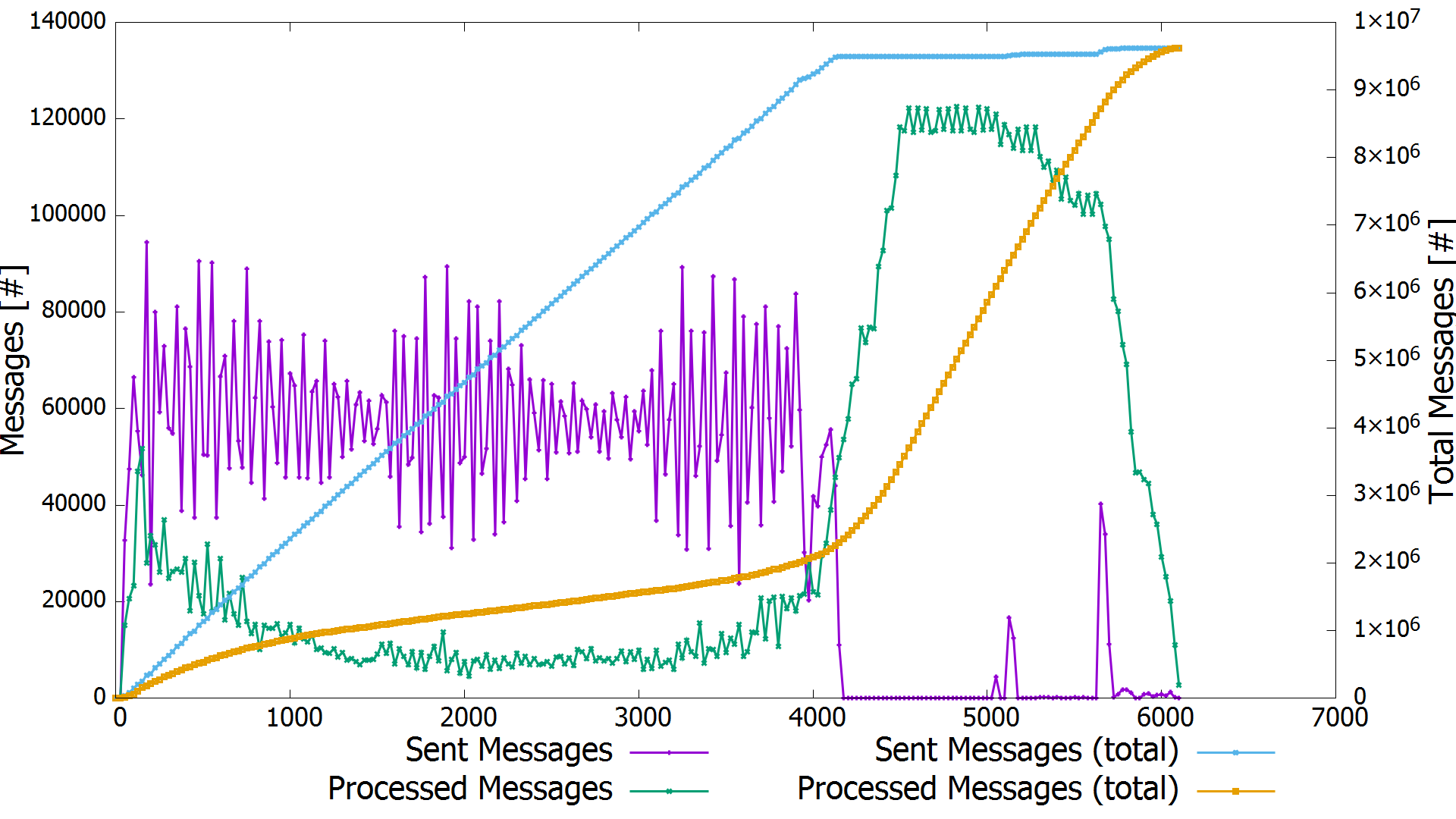}}
\\	
	\subfigure[b][\emph{Lookup-kWay} design]{\label{fig:detailed_kway}\includegraphics[width=0.48\textwidth]{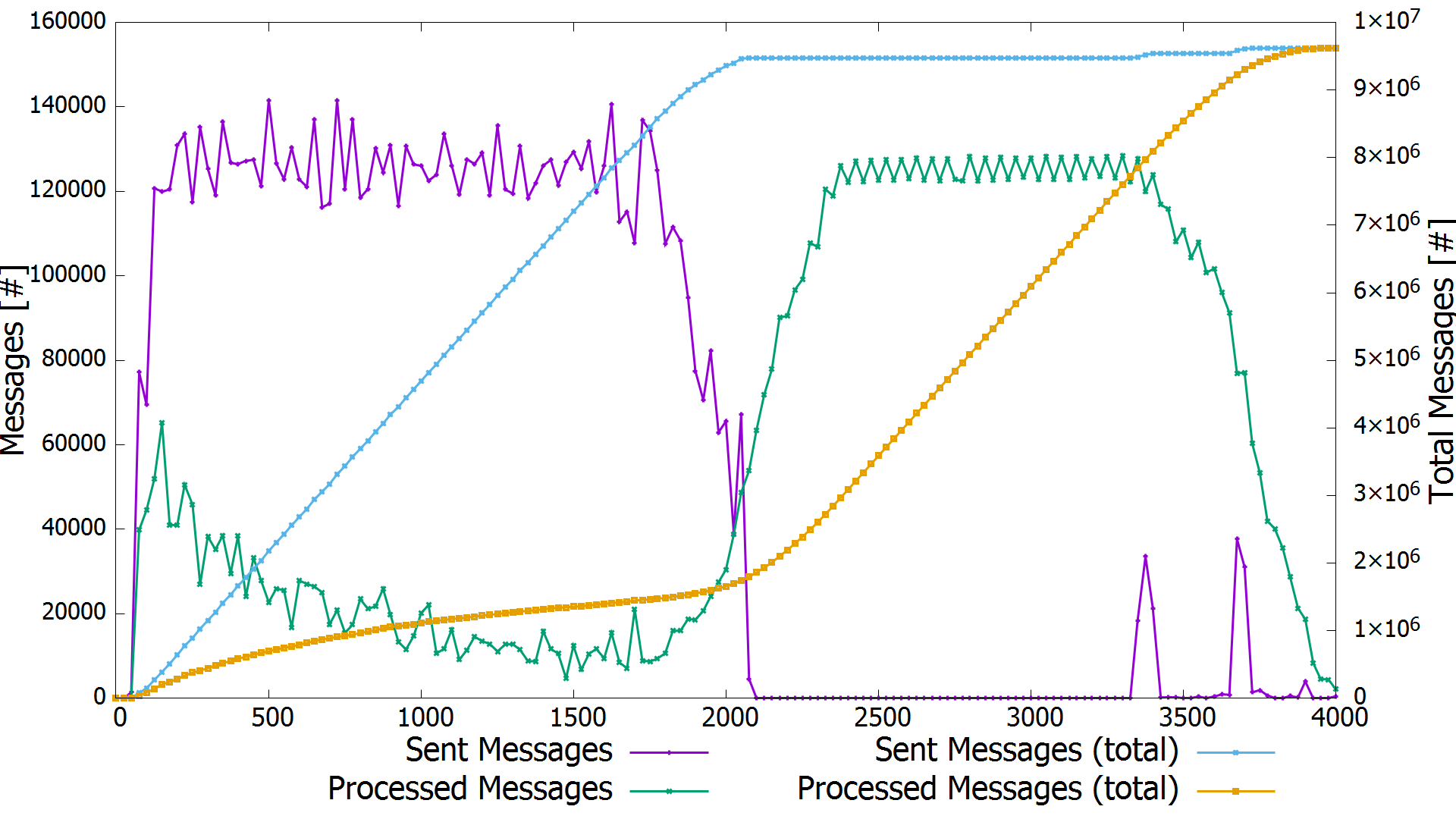}}
	\hfill   
	\subfigure[b][\emph{Hybrid} design]{\label{fig:detailed_rv}\includegraphics[width=0.48\textwidth]{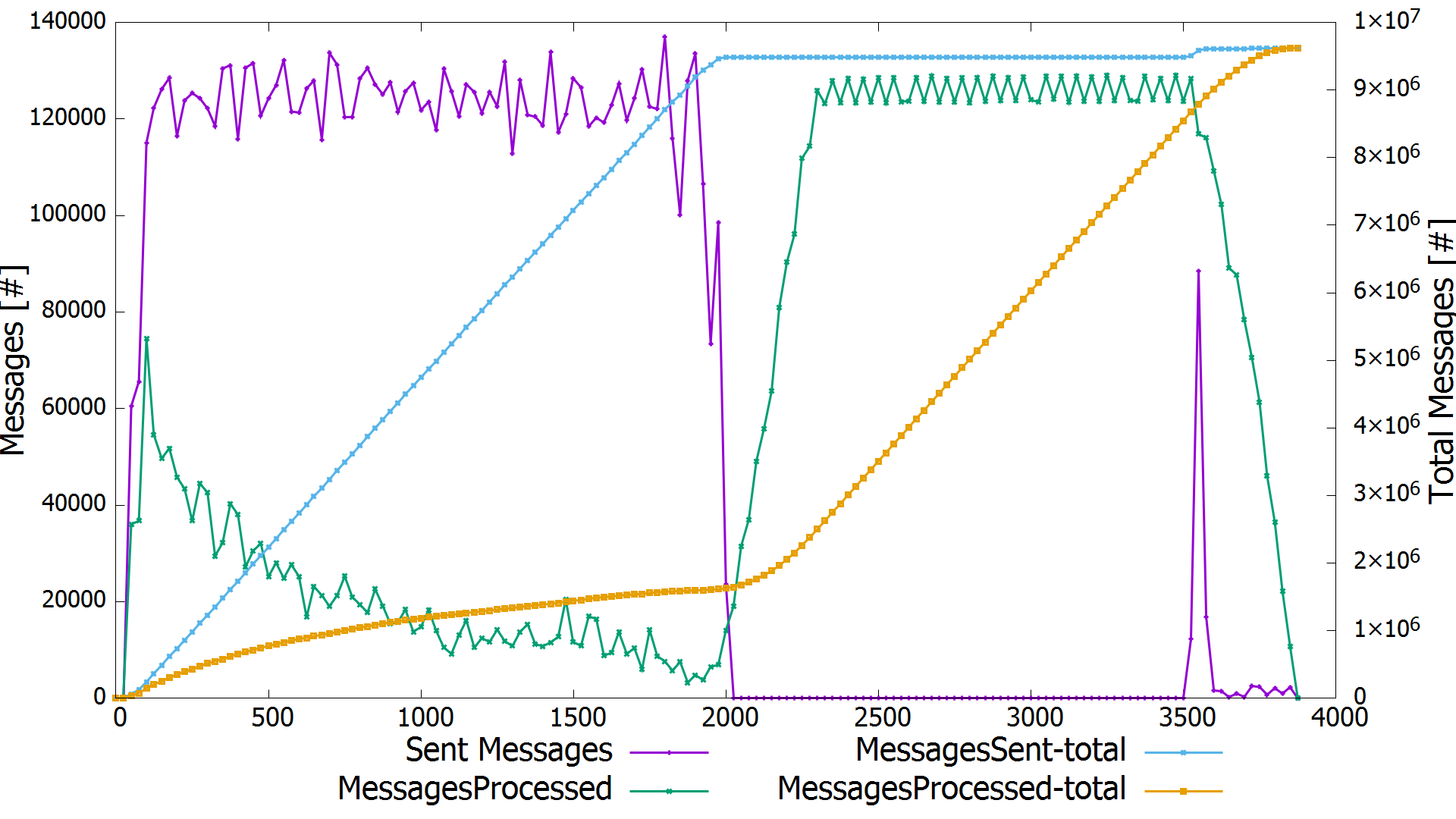}}
	\caption{Messaging behavior for the \emph{V} query on the Biblio graph.}
	\label{fig:parallelExecution}
\end{figure}
Figure~\ref{fig:routingTime_nored} shows the differences between the times spent in the routing table, for the different combinations.
As anticipated, the \emph{compute design} leads to the least time spent in the routing table, because it only computes the target partitions using a hash function.
For the \emph{lookup design}, we measured two different \emph{partitioning strategies}.
The first is the locality-agnostic partitioning as it is employed in the \emph{compute design} (lookup-hash) to show the structural overhead which results from the size of the routing table.
The second is the locality-aware partitioning (lookup-kWay). 
As shown, the lookup-based routing table requires the most time in the routing table, because of the high number of main memory accesses.
Our \emph{hybrid design} clearly outperforms the lookup-based routing table and requires only a little more time compared to the \emph{compute design}.

Figure~\ref{fig:queryTime_nored} visualizes the corresponding query runtimes.
We observe that both -- the \emph{lookup-kWay design} and our \emph{hybrid design} -- outperform the designs that employ a locality-agnostic \emph{partitioning strategy}.
Hence, for this specific setup of graph and query pattern, the query runtime is only affected by the \emph{partitioning strategy}.
However, the huge routing time differences (cf., Figure~\ref{fig:routingTime_nored}) are not mirrored in the \emph{hybrid} and \emph{lookup-kWay design} as well as in both locality-agnostic designs.
This phenomenon can be explained with the matching vertices per edge predicate request, because most of the messages are generated in the early phase of the query execution.
This leads to a small overall impact of the \emph{routing table} on the query runtime.
To emphasize the importance of this load bias, we visualized the message passing behavior for the different designs in  Figure~\ref{fig:parallelExecution}.
As shown for all of the designs, the system is predominately producing messages within the first half of the query execution, because the first operators in the QEP sequence are likely to produce a significant amount of matches. 
We observe the same effect for the locality-agnostic designs.


\subsection{Avoiding Broadcasts with Redundancy}
\label{sec:evaluationBroadcasts}

\begin{figure}[t]
	\centering
	\includegraphics[width=\textwidth]{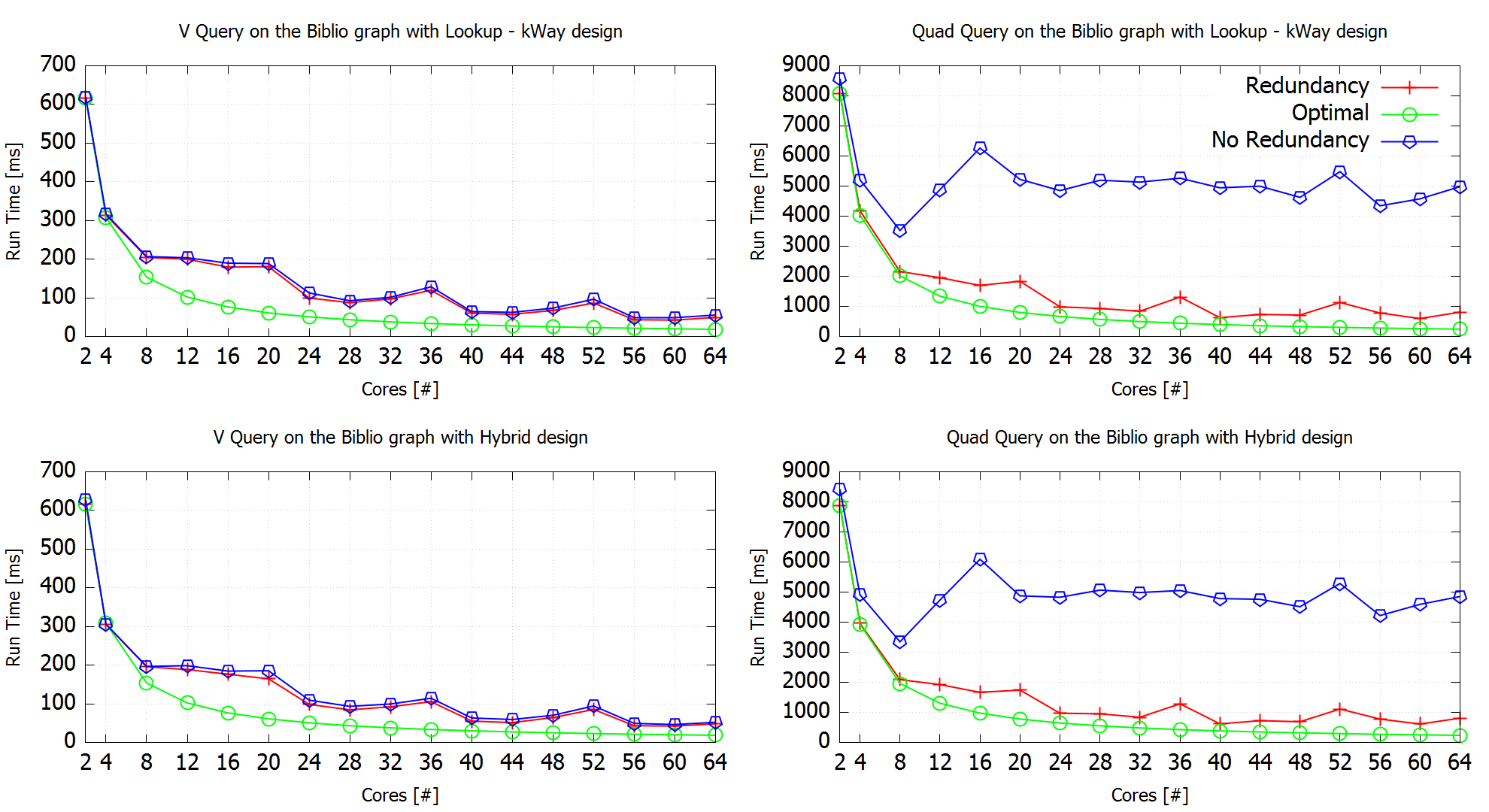}
	\caption{Impact of Redundancy, both queries on the Biblio graph.}
	\label{fig:redundancy_Biblio}
\end{figure}

To prove the influence of broadcasts on the query performance, we use all graphs from Figure~\ref{fig:GraphMeta} and the query patterns \emph{V} and \emph{Quad}.
Because both the \emph{lookup-kWay} and \emph{hybrid design} routing tables outperform the \emph{compute design} and \emph{lookup-hash} routing tables as shown in Figure~\ref{fig:queryTime_nored}, we limit the presentation of further experiments to these two for better readability.
Figure~\ref{fig:redundancy_Biblio} shows the query performance for both query patterns on the Biblio graph.
For this experiment, we varied the number of active workers and ran the queries first without using redundant information and second with redundancy in the data.
The green line show the calculated ideal runtime, based on the performance for two workers.

\begin{figure}[t]
	\centering
	\includegraphics[width=\textwidth]{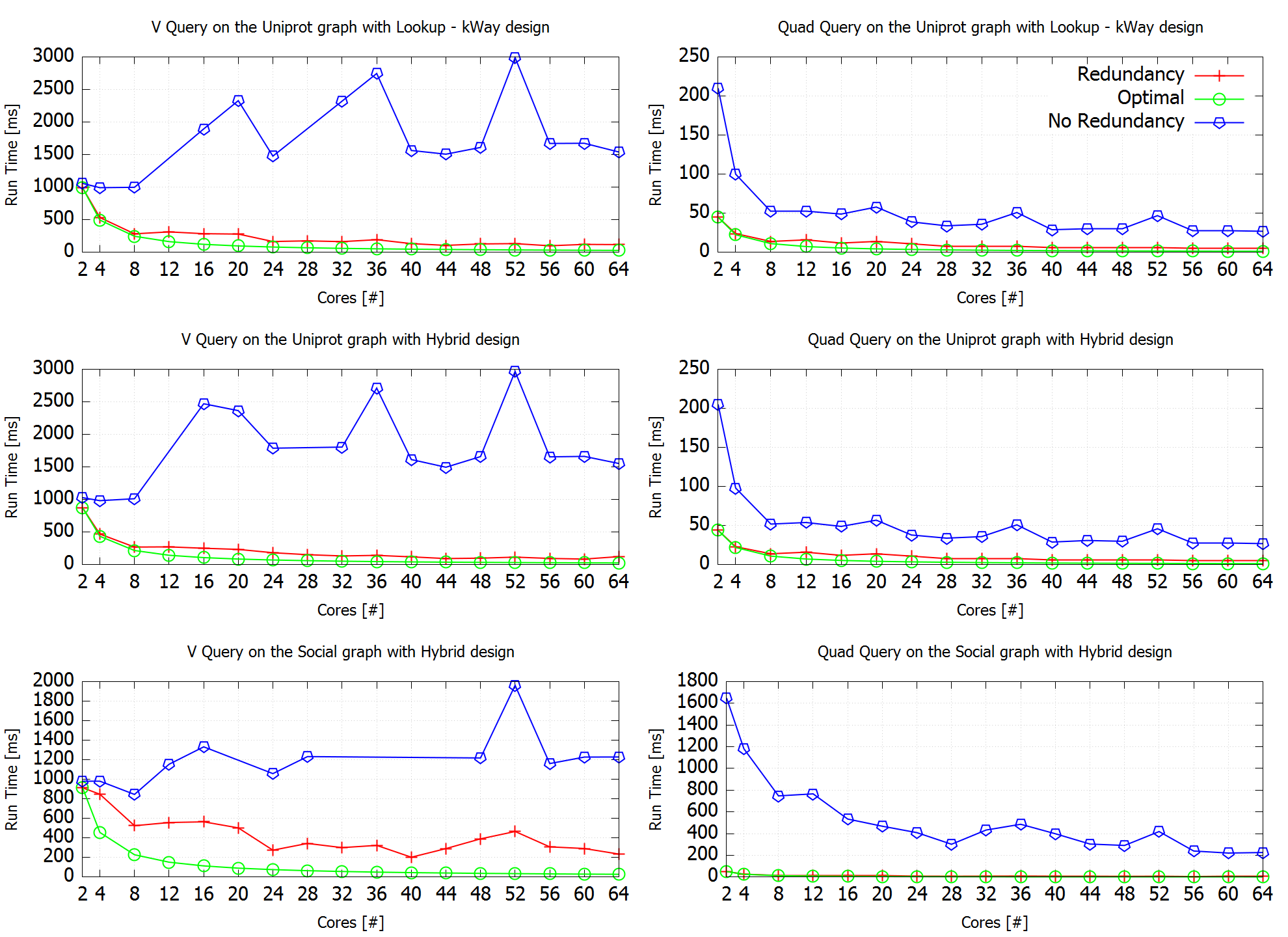}
	\caption{Impact of Redundancy, both queries on the Uniport and Social graph}
	\label{fig:redundancy_UniprotSocial}
\end{figure}

In this figure, we see that there is a difference in the benefit for the employment of redundancy.
The \emph{V} query pattern is depicted at the left hand side of the figure. We see that the usage of redundancy only shows a minimal positive impact on the query performance.
However, the \emph{Quad} query pattern highly benefits from it.
This is because of the number of vertices in the data graph, which match a query vertex within an operator, which sends out a broadcast.
For the \emph{V} query pattern on the Biblio graph, there are only very few matches for the operators which reside later in the chain of the QEP (c.f. Figure~\ref{fig:VQuery}).
This means, that the system does not send many broadcasts in these steps and thus the broadcasting overhead is only marginal.
The similarity between the measurements of the \emph{hybrid design} and the \emph{lookup-kWay} measurements is explained with the fact, that the \emph{hybrid design} uses a k-Way partitioning underneath.

On the right hand side of Figure~\ref{fig:redundancy_Biblio}, we see that the \emph{Quad} query pattern leads to a totally different system behavior than the \emph{V} query pattern.
For the \emph{Quad} query pattern, we can observe that the broadcasting operators within the QEP find many matching vertices.
This leads to a significant amount of broadcasting messages in the system and inhibits its scalability. 

The dents in the plots are explained with the activation of hyper threads and the usage of an additional socket.
Because the hyper threads don't provide the same compute power as a physical core, switching on the SMT sibling of an active worker can even lead to a small performance decrease for compute intensive tasks.
By switching on a new processor with only a portion of its threads, we see a little performance decrease because of the NUMA effect.
As soon as all physical threads of a processor are used as worker threads, the system performance reaches a local minimum.
However, we can see for both query patterns on the Biblio graph, that our system scales well with the number of employed workers.

Figure~\ref{fig:redundancy_UniprotSocial} shows the results for the same experiments but with the Uniprot and the Social graph. 
For both graphs, we found that both query patterns behave contrary to the Biblio graph.
The \emph{V} query pattern finds many matches during the broadcasting operators and the \emph{Quad} query pattern produces far less broadcasts without the redundant implementation.
However, both queries can benefit from redundant data and the resulting unicast-only execution.
That redundancy can also achieve a tremendous performance gain is shown by the \emph{Quad} query pattern on the Social graph on the bottom right side of Figure~\ref{fig:redundancy_UniprotSocial}.
The redundant execution achieves a query runtime of \SI{6}{ms} with \SI{64} workers, which is faster than the non-redundant execution by a factor of \SI{37.7}.
From the Figures~\ref{fig:redundancy_Biblio}~and~\ref{fig:redundancy_UniprotSocial} we can derive that it is always beneficial for the query performance, to leverage redundancy in terms of partitioning.
Dependent on the underlying graph data, a query pattern can lead to many broadcasts or almost none. 
However, the query performance does not suffer from using redundancy and it allows the system to scale up well for most scenearios.

\begin{figure}[t]
	\centering
	\subfigure[b][Routing time]{\label{fig:routingTime_red}\includegraphics[width=0.48\textwidth]{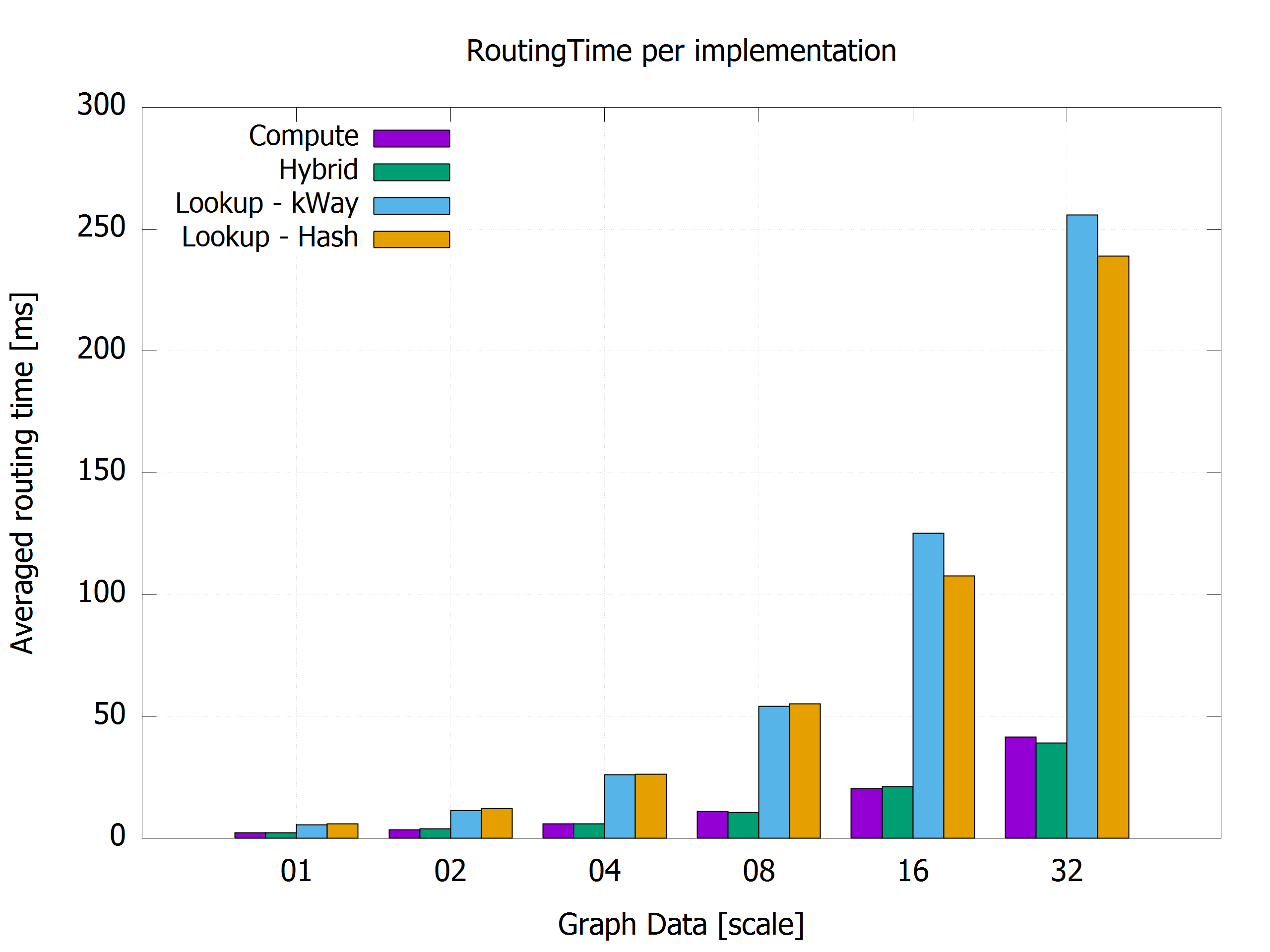}}
	\hfill   
	\subfigure[b][Query runtime]{\label{fig:queryTime_red}\includegraphics[width=0.48\textwidth]{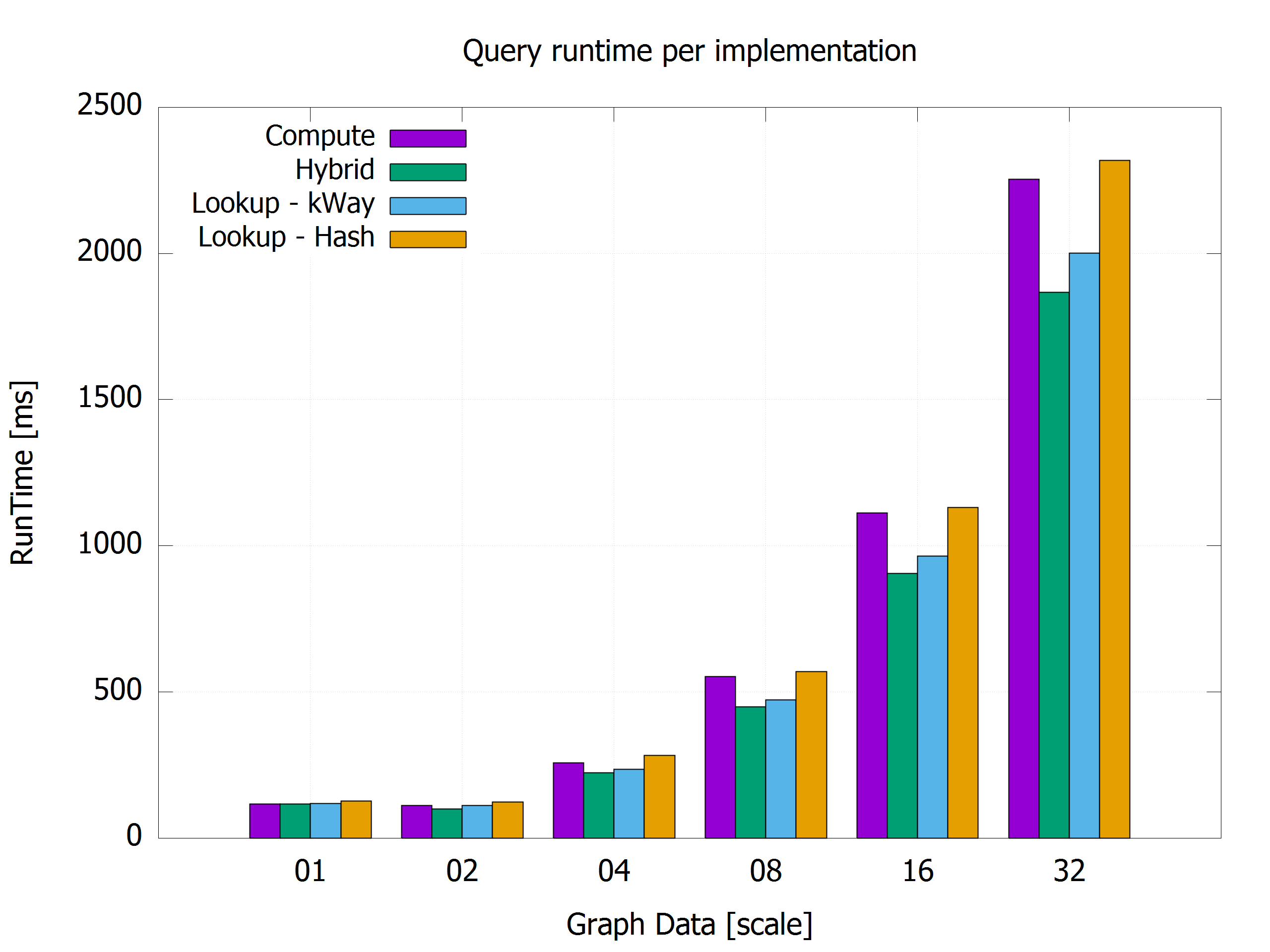}}
	\caption{\emph{V} query on the Biblio graph, scaled up to factor 32.}
	\label{fig:routingVsQuerytime_red}
\end{figure}

\subsection{Combining Redundancy and Routing Table Optimizations}
We haven proven the positive effects of our \emph{hybrid design} and the usage of redundancy on the query performance for multiple graphs.
For the experiment shown in Figure~\ref{fig:routingVsQuerytime_red}, we have combined both optimizations to show that their effects can add up and improve the system performance even further. 

Figure~\ref{fig:routingTime_red} shows the average time spent in the routing table per worker and Figure~\ref{fig:queryTime_red} shows the runtime for the \emph{V} query on all graph scale factors.
We can see, that both the \emph{compute design} and \emph{hybrid design} implementations still outperform the \emph{lookup} versions in the routing table performance.
Depending on the size of the queried graph, the query performance difference in Figure~\ref{fig:queryTime_red} between the \emph{compute design} and \emph{lookup-hash design} range from \SI{2}{\%} up to \SI{10}{\%} and we can achieve a speedup of factor 2.6.
With our \emph{hybrid design}, which uses a locality-aware graph partitioning with a small routing table and redundant data execution we are able to improve on the query performance of the non-optimized \emph{compute design} by a factor of 3.2.
By comparing Figure~\ref{fig:routingVsQuerytime_red} with Figure~\ref{fig:routingVsQuerytime_nored} we still find the same phenomenon, that the routing time does not linearly add up to the query runtime. 
We identify the same messaging behavior as reason for this effect like we mentioned in Section~\ref{sec:evaluationRouting} regarding Figure~\ref{fig:routingVsQuerytime_nored}.

\/*
\subsection{Evaluation on a Large-Scale Multiprocessor System}
\label{sec:largeEval}
Our large multiprocessor system is an SGI UV 3000 with 64 sockets each equipped with an Intel Xeon CPU E5-4655 v3 and a total of 8 TB main memory. We conducted the same experiments as for Section~\ref{sec:mediumEval} and used gMark to scale up all graphs from Figure~\ref{fig:GraphMeta} by a factor of $10$ while preserving all other graph properties.

The scalability experiment is presented in Figure~\ref{fig:sgiScaling}.
Because we increased the data by a factor of 10, we started our measurements with an initial worker count of 120, which run on 10 processors, since each processor features 12 hardware threads.
We see in both Figures~\ref{fig:scaling_biblio_sgi_kway}~and~\ref{fig:scaling_biblio_sgi_rv}, that the query runtime decreases with an increasing amount of processors, although the scaling factor is very small.
This can be explained by the problems complexity.
The query consists of only four edge predicate requests and despite of the increased data set size, there are not many qualifying vertices for the query pattern.
This means that only a limited amount of partitions is targeted by the query. 
Following Amdahl's Law, the system can only scale up to a certain threshold, until the parallel portion of the workload can not be divided any further.
Thus adding more workers only inhibits the system performance rather than increasing it.
However, the spikes in Figure~\ref{fig:scaling_biblio_sgi_rv} show that certain local optimal partitionings exist, which may result in a better runtime than using the maximum amount of workers.

\begin{figure}[t]
	\centering
	\subfigure[b][System scaling for \emph{V} query on the Biblio graph with k-Way partitioning]{\label{fig:scaling_biblio_sgi_kway}\includegraphics[width=0.48\textwidth]{KWay_scalingBiblio_V_SGI.png}}
	\hfill   
	\subfigure[b][System scaling for \emph{V} query on the Biblio graph with ranged partitioning]{\label{fig:scaling_biblio_sgi_rv}\includegraphics[width=0.48\textwidth]{RangedVertices_scalingBiblio_V_SGI.png}}
	\caption{System Scaling behavior on the SGI UV 3000}
	\label{fig:sgiScaling}
\end{figure}
*/

\/*
\begin{itemize}
\item Hypothesis: RoutingTable heavily influences QueryPerformance
\item Explanation Setup: Machine, Graph, Workload(s)
\item Experiments Query Runtime for Hash, Hashtable, Hash with RB Tree -- same Partition Quality, but other access strategy
\end{itemize}
*/

\section{Related Work}
\label{sec:relatedWork}
Graph analytics is a widely studied field, as the survey from McCune et al. \cite{McCune:2015:TLV:2830539.2818185} shows.
Many systems leverage the increased compute performance of scale-up or scale-out systems to compute graph metrics like PageRank and the counting of triangles \cite{Gonzalez2012,DBLP:journals/corr/abs-1107-0922,DBLP:journals/pvldb/SeoPSL13}, Single Source Shortest Path \cite{Malewicz:2010:PSL:1807167.1807184,DBLP:conf/ppopp/ShunB13,DBLP:journals/pvldb/SeoPSL13} or Connected Components \cite{gonzalez2014graphx,DBLP:conf/ppopp/ShunB13,DBLP:journals/pvldb/SeoPSL13}.
Many of the available systems are inspired by the Bulk Synchrones Processing Model \cite{Valiant:1990:BMP:79173.79181}, which features local processing phases which are synchronized by a global superstep.
A general implementation is the Gather-Apply-Scatter paradigm, as described in \cite{Gonzalez2012}.
Despite working on NUMA systems, these processing engines are globally synchronized and lack the scalability of a lock-free architecture. 
We improve this issue by leveraging a high throughput message passing layer for asynchronous communication between the worker threads.
However, in contrast to the systems mentioned above, we are calculating graph pattern matching and not graph metrics, like for instance GraphLab which is the only asynchronous graph processing engine according to~\cite{McCune:2015:TLV:2830539.2818185}.

The authors of Polymer \cite{DBLP:conf/ppopp/ZhangCC15} show that is important to consider NUMA effects for graph processing.
One of their findings is, that sequential remote read operations are more performant than random local reads and reason that the data layout and a well designed communication is crucial for high performance.
Like the authors, we found that an optimized communication is a key characteristic for high performance.
In contrast to the Polymer system, our processing engine does not directly access remote memory but rather sends a message to a worker, which can then process its local data.

\section{Conclusions and Future Work}
\label{sec:conclusion}

In this paper, we showed that the performance of graph pattern matching on a multiprocessor system is determined by the communication behavior, the employed routing table design and the partitioning strategy.
We could show that the design of the routing table has a remarkable impact on the systems query performance.
The system performance is not only determined by the design of the routing table, but also by the quality of the underlying partitions (c.f. Figure~\ref{fig:routingVsQuerytime_red}).
Our \emph{Hybrid} design implementation allows the system to leverage both the advantages from a \emph{Compute} design and a \emph{Lookup} design.
Because of an intermediate step, the underlying graph partitioning algorithm is interchangable and can thus be adapted to specific partitioning requirements.
Furthermore we could show that broadcasts can impose a major performance bottleneck for the graph pattern matching process (c.f. right hand side of Figure~\ref{fig:redundancy_Biblio} and left hand side of Figure~\ref{fig:redundancy_UniprotSocial}).
This issue was mitigated by introducing redundancy in the system.
However, by additionally storing the incoming edges for each vertex we double the memory footprint of the graph.
This penalty is not only paid for the data storage, but also for the routing table, if a \emph{Lookup} or \emph{Hybrid} design is used.
However, our \emph{Hybrid} design scales directly with the total number of data partitions in the system and introducing redundancy for our \emph{Hybrid} table does only lead to a minimal memory penalty.

\section{Acknowledgments}
This work is partly funded by the German Research Foundation (DFG) in the CRC 912 (HAEC).

%
\bibliographystyle{abbrv}
\bibliography{references}  

\end{document}